\font\bbb=msbm10 \font\bbs=msbm7                                   

\overfullrule=0pt

\input epsf.tex
\input pstricks  
\input color
\definecolor{Green}{rgb}{0.25,0.75,0.25}
\definecolor{review}{rgb}{0.25,0.75,0.25}
\definecolor{exercise}{rgb}{1,0,0}
\definecolor{lightgray}{rgb}{0.5,0.5,0.5}

\def\C{\hbox{\bbb C}}

\def\Z{\hbox{\bbb Z}} \def\sZ{\hbox{\bbs Z}} 
\def\Tr{{\rm Tr}}
\def\Pr{{\rm Pr}}
\def\span{{\rm span}}
\def\Re{{\rm Re}}

\def\CACM{{\sl Commun.\ ACM\/}}

\def\IC{{\sl Inform.\ Control}}

\def\IEEETIT{{\sl IEEE Trans.\ Inform.\ Theory}}

\def\JACM{{\sl J. ACM\/}}

\def\JMA{{\sl J. Multiv.\ Anal.}}

\def\JMO{{\sl J. Mod.\ Optics\/}}

\def\ML{{\sl Machine Learning}}

\def\PIEEEL{{\sl Proc.\ IEEE\/} ({\sl Lett.\/})}

\def\PRA{{\sl Phys.\ Rev.\ A\/}}

\def\PRL{{\sl Phys.\ Rev.\ Lett.}}
\def\PRSLA{{\sl Proc.\ Roy.\ Soc.\ Lond.\ A\/}}

\def\QIP{{\sl Quantum Inform.\ Processing}}

\def\SIAMJC{{\sl SIAM J. Comput.}}

\def\dajm{\hbox{D. A. Meyer}}

\def\jamie{\hbox{J. Pommersheim}}
\def\dj{\hbox{\dajm\ and \jamie}}

\def\bv{\hbox{E. Bernstein and U. Vazirani}}

\def\emv{\hbox{Y. C. Eldar, A. Megretski and G. C. Verghese}}

\def\grover{\hbox{L. K. Grover}}

\def\shor{\hbox{P. W. Shor}}
\def\simon{\hbox{D. R. Simon}}

\def\vandam{\hbox{W. van Dam}}

\def\hfb{\hfil\break}

\catcode`@=11
\newskip\ttglue

   \font\ninerm=cmr9    \font\eightrm=cmr8   \font\sixrm=cmr6
  \font\ninebf=cmbx9   \font\eightbf=cmbx8  \font\sixbf=cmbx6
  \font\nineit=cmti9   \font\eightit=cmti8  
  \font\ninesl=cmsl9   \font\eightsl=cmsl8  
  \font\ninemi=cmmi9   \font\eightmi=cmmi8  \font\sixmi=cmmi6

\font\bigten=cmr10 scaled\magstep2 

\def\ninepoint{\def\rm{\fam0\ninerm}%
  \textfont0=\ninerm \scriptfont0=\sixrm
  \textfont1=\ninemi \scriptfont1=\sixmi
  \textfont\itfam=\nineit  \def\it{\fam\itfam\nineit}%
  \textfont\slfam=\ninesl  \def\sl{\fam\slfam\ninesl}%
  \textfont\bffam=\ninebf  \scriptfont\bffam=\sixbf
    \def\bf{\fam\bffam\ninebf}%
  \tt \ttglue=.5em plus.25em minus.15em
  \normalbaselineskip=11pt
  \setbox\strutbox=\hbox{\vrule height8pt depth3pt width0pt}%
  \normalbaselines\rm}

\def\eightpoint{\def\rm{\fam0\eightrm}%
  \textfont0=\eightrm \scriptfont0=\sixrm
  \textfont1=\eightmi \scriptfont1=\sixmi
  \textfont\itfam=\eightit  \def\it{\fam\itfam\eightit}%
  \textfont\slfam=\eightsl  \def\sl{\fam\slfam\eightsl}%
  \textfont\bffam=\eightbf  \scriptfont\bffam=\sixbf
    \def\bf{\fam\bffam\eightbf}%
  \tt \ttglue=.5em plus.25em minus.15em
  \normalbaselineskip=9pt
  \setbox\strutbox=\hbox{\vrule height7pt depth2pt width0pt}%
  \normalbaselines\rm}

\def\sfootnote#1{\edef\@sf{\spacefactor\the\spacefactor}#1\@sf
      \insert\footins\bgroup\eightpoint
      \interlinepenalty100 \let\par=\endgraf
        \leftskip=0pt \rightskip=0pt
        \splittopskip=10pt plus 1pt minus 1pt \floatingpenalty=20000
        \parskip=0pt\smallskip\item{#1}\bgroup\strut\aftergroup\@foot\let\next}
\skip\footins=12pt plus 2pt minus 2pt
\dimen\footins=30pc

\def\ie{{\it i.e.}}
\def\eg{{\it e.g.}}

\def\mod{{\rm mod}\ }
\def\calbv{{\cal BV}}

\def\calo{{\cal O}}

\def\psucc{p_{\rm{success}}}

\def\Lemma{L{\eightpoint EMMA}}

\def\Theorem{T{\eightpoint HEOREM}}

\def\Corollary{C{\eightpoint OROLLARY}}
\def\Definition{D{\eightpoint EFINITION}}

\def\Question{Q{\eightpoint UESTION}}
\def\Proposition{P{\eightpoint ROPOSITION}}
\def\SOD{S{\eightpoint YMMETRIC} O{\eightpoint RACLE} D{\eightpoint ISCRIMINATION}}
\def\GM{G{\eightpoint ROUP} M{\eightpoint ULTIPLICATION}}
\def\HCEP{H{\eightpoint IDDEN} C{\eightpoint ONJUGATING} E{\eightpoint LEMENT} P{\eightpoint ROBLEM}}
\def\PARITY{P{\eightpoint ARITY}}
\def\endproof{\vrule height6pt width4pt depth2pt}

\def\and{{\eightpoint AND}}

\def\Kholevo{1}
\def\YKL{2}
\def\Helstrom{3}
\def\HW{4}
\def\SKIH{5}
\def\EldarForney{6}
\def\EMV{7}
\def\MeyerPommersheim{8}
\def\Deutsch{9}
\def\DeutschJozsa{10}
\def\MeyerPommersheimB{12}
\def\BBCMdW{11}
\def\vanDam{13}
\def\BV{15}
\def\Bhatia{14}
\def\Frumkin{16}
\def\Roth{17}
\def\Formanek{18}
\def\Roichman{19}

\magnification=1000

\input epsf.tex

\dimen0=\hsize \divide\dimen0 by 13 \dimendef\chasm=0
\dimen1=\chasm \multiply\dimen1 by  6 \dimendef\halfwidth=1
\dimen2=\chasm \multiply\dimen2 by  7 \dimendef\secondstart=2
\dimen3=\chasm \divide\dimen3 by 2 \dimendef\quarter=3
\dimen4=\quarter \multiply\dimen4 by 9 \dimendef\twopointtwofivein=4
\dimen5=\chasm \multiply\dimen5 by 3 \dimendef\onepointfivein=5
\dimen6=\chasm \multiply\dimen6 by 7 \dimendef\threepointfivein=6
\dimen7=\hsize \advance\dimen7 by -\chasm \dimendef\usewidth=7
\dimen8=\chasm \multiply\dimen8 by 4 \dimendef\thirdwidth=8
\dimen9=\usewidth \divide\dimen9 by 2 \dimendef\halfwidth=9
\dimen10=\usewidth \divide\dimen10 by 3  
                   \multiply\dimen10 by 2 \dimendef\twothirdswidth=10

\parskip=0pt\parindent=0pt

\line{\hfil 18 March 2015}
\vfill
\centerline{\bf\bigten DISTINGUISHING SYMMETRIC QUANTUM ORACLES}
\smallskip
\centerline{\bf\bigten AND QUANTUM GROUP MULTIPLICATION}
\bigskip\bigskip
\centerline{\bf Orest Bucicovschi$^*$, Daniel Copeland$^*$, David A. Meyer$^*$}
\centerline{\bf and James Pommersheim$^{*,\dagger}$}
\bigskip 
\centerline{\sl $^*$Project in Geometry and Physics,
                Department of Mathematics}
\centerline{\sl University of California/San Diego,
                La Jolla, CA 92093-0112}                
\smallskip
\centerline{\sl $^{\dagger}$Department of Mathematics}
\centerline{\sl Reed College, Portland, OR 97202-8199}
\smallskip
\centerline{{\tt obucicov@math.ucsd.edu}, {\tt drcopela@math.ucsd.edu}, 
            {\tt dmeyer@math.ucsd.edu}}
\centerline{and {\tt jamie@reed.edu}}            
\smallskip

\vfill
\centerline{ABSTRACT}
\bigskip
\noindent Given a unitary representation of a finite group on a 
finite-dimensional Hilbert space, we show how to find a state whose 
translates under the group are distinguishable with the highest 
probability.  We apply this to several quantum oracle problems, 
including the \GM\ problem, in which the product of an ordered 
$n$-tuple of group elements is to be determined by querying elements 
of the tuple.  For any finite group $G$, we give an algorithm to find 
the product of two elements of $G$ with a single quantum query with 
probability $2/|G|$.  This generalizes Deutsch's Algorithm from $\Z_2$ 
to an arbitrary finite group.  We further prove that this algorithm is 
optimal.  We also introduce the \HCEP, in which the oracle acts by 
conjugating by an unknown element of the group.  We show that for many 
groups, including dihedral and symmetric groups, the unknown element 
can be determined with probability $1$ using a single quantum query.

\bigskip\bigskip
\noindent 2008 Physics and Astronomy Classification Scheme:
                   03.67.Ac. 

\noindent 2000 American Mathematical Society Subject Classification:
                   81P68,    
                   68Q32,    
                   05B20.    

\smallskip
\global\setbox1=\hbox{Key Words:\enspace}
\parindent=\wd1
\noindent Key Words: quantum state discrimination, quantum algorithms.

\vfill
\eject

\headline{\ninepoint\it Symmetric oracles and group multiplication 
                   \hfill Bucicovschi, Copeland, Meyer \& Pommersheim}

\parskip=10pt
\parindent=0pt

\noindent{\bf 1.  Introduction}

Given a finite set of states in a finite dimensional Hilbert space, 
one may wish to find an optimal measurement, \ie, a measurement that 
distinguishes the states of the set with maximal probability.  Indeed, 
this situation arises near the end of many quantum algorithms, when 
the system is in one of a finite number of states depending on some 
unknown or hidden information, and one wishes to determine this 
information with maximal probability by means of a quantum 
measurement.

There is an extensive literature on such quantum detection problems 
[\Kholevo
--\EMV], and in many cases the optimal measurement is known.  Of 
particular interest is the case in which the collection of states is 
{\sl geometrically uniform}, meaning that there is a finite group of 
unitary maps that acts transitively on the collection.  When these
states are pure, the least-squares measurement (or square-root 
measurement) is known to distinguish them optimally [\EMV]. 

One way to obtain a geometrically uniform collection of states is to 
begin with a finite group of unitary operators, and consider the orbit 
of a given initial state $|\psi\rangle$ under the group.  The various 
detection theorems then tell us how optimally to distinguish the 
elements of this orbit.  Suppose that we are also given the freedom to 
choose the initial state.  How should we choose $|\psi\rangle$ so that 
images of $|\psi\rangle$ under the group action are maximally 
distinguishable when the optimal measurement is made?  This  question, 
which we call \SOD, is the first question that we address in this 
paper.

We can frame this question more precisely:  Let $G$ be a finite group 
and let $\Theta:G\rightarrow GL(V)$ be a unitary representation of $G$ 
on a finite-dimensional Hilbert space $V$.  Assume that an unknown 
element $g\in G$ is chosen uniformly at random and we wish to find 
$g$.  To do this, we are allowed to chose any initial state 
$|\psi\rangle\in V$, apply the unitary map $\Theta(g)$ to create the 
state $|\psi_g\rangle := \Theta(g) |\psi\rangle$, and then apply an 
arbitrary POVM $\{M_h\}_{h\in G}$.  We wish to choose $|\psi\rangle$ 
and $\{M_h\}_{h\in G}$ to maximize the probability that the unknown 
$g\in G$ is measured.  We may view this as an oracle problem:
determine the unknown element $g\in G$ using a single query to an 
oracle that acts on $V$ by ${\cal O}_g = \Theta(g)$.
 
Since for any $|\psi\rangle\in V$, the collection 
$\{\Theta(g)|\psi\rangle\}$ is geometrically uniform, the optimal 
measurement for distinguishing these states consists of rank 1 
projections $M_h=|\mu_h\rangle\langle \mu_h |$ that sum to the 
identity [\EMV].  For such a measurement, the probability of measuring 
$h$ given that $g$ is the chosen element is given by
$$
\Pr(h|g) = \Tr(M_h |\psi_g\rangle\langle \psi_g |) 
         = |\langle\mu_h| \psi_g\rangle |^2.
$$
Thus our question can be phrased as follows: 
 
\noindent\SOD.  Given a representation $\Theta:G\rightarrow GL(V)$ of 
a finite group $G$ on a finite-dimensional Hilbert space $V$, 
determine a state $|\psi\rangle\in V$ and collection 
$\{\mu_h\}_{h\in G}$ of vectors in $V$ such that 
$$
\sum_{h\in G} |\mu_h\rangle\langle \mu_h| = I_V
$$
and so that the success probability
$$
\psucc = {{1}\over{|G|}}\sum_{g\in G} |\langle\mu_g| \psi_g\rangle |^2
$$
is maximized.
 
As a first observation, we note that the success probability above is 
limited by the dimension of the span of the translates of 
$|\psi\rangle$.  Indeed, if we have any $n$ vectors in a Hilbert space 
which span a subspace of dimension $k$, then a quantum measurement can 
distinguish among these vectors with probability at most $k/n$ 
({\it cf}.~[\MeyerPommersheim] and Lemma 1 below). 
 
For a given representation $\Theta$, we let $d_{\Theta}$ denote the 
maximum over all choices of initial state $|\psi\rangle$ of the 
dimension of the span of the orbit of $|\psi\rangle$ under the action 
of $G$.  That is,
$$
d_{\Theta} 
 = \max_{|\psi\rangle} \dim \span\{\Theta(g)|\psi\rangle\mid g\in G\}.
$$
The number $d_{\Theta}$ is the maximal dimension of a cyclic subspace 
of $V$ and can easily be expressed in terms of the decomposition of $V$ 
into irreducible representations (see Lemma 2).  

It follows that for any representation $\Theta$, we have
$$
\psucc\leq {d_{\Theta}\over |G|}.
$$
One may then ask if for any representation, one can find a state that 
achieves this dimension bound.  An affirmative answer is given by the 
following theorem.

\noindent\Theorem~1.  {\sl Given a representation 
$\Theta:G\rightarrow GL(V)$ of a finite group $G$ on a 
finite-dimensional Hilbert space $V$, there exists a state 
$|\psi\rangle\in V$ for which {\rm\SOD} can be solved with probability 
$$
\psucc = {d_{\Theta}\over |G|}.
$$
}
 
The construction of an initial state $|\psi\rangle$ that achieves this 
dimension bound is given during the course of the proof of this 
theorem (see Propositions 1, 2 and 3 in \S2).  In fact, in that 
section we completely characterize the initial states that achieve the 
dimension bound.

The second question we address in this paper has its roots in one of
the first quantum algorithms, Deutsch's algorithm, which solved the 
problem of finding the mod 2 sum of two bits $(b_0, b_1)$ using a 
single quantum query [\Deutsch,\DeutschJozsa].  For the purpose of 
this query, we have a Hilbert space $V=\C^2\otimes\C^2$, where the 
first tensor factor is the query register and the second the response 
register.  We are given access to an oracle ${\cal O}_{(b_0,b_1)}$ 
whose action on $V$ is given by
$$
{\cal O}_{(b_0,b_1)}: |j,r\rangle \mapsto |j,b_j+r \rangle, 
$$
where the addition in the response register takes place in the group 
$G=\Z_2$.  This problem admits a probability 1 solution with a single 
quantum query.  We may also ask for the sum of more than two bits; 
this is the \PARITY\ problem studied in [\BBCMdW] and elsewhere.

One may naturally generalize this problem to any finite group $G$,  
abelian or nonabelian.  Given an $m$-tuple $(g_0,g_1,\ldots,g_{m-1})$ 
of elements of $G$, we wish to determine the product 
$g_0 g_1 \cdots g_{m-1}$ of the group elements in the given order.  To 
set this up as a quantum oracle problem, we take a query register 
$\C^m$ with basis $\{|j\rangle\mid j=0,\ldots,m-1\}$ and response 
register $\C^G$ with basis $\{ |g\rangle\mid g\in G\}$ labeled by the 
elements of $G$.  When the $j^{\rm th}$ element is queried, the oracle 
multiplies $g_j$ into the response register.  Formally, we have:

\noindent\GM.  Let $G$ be a finite group.  Suppose that an unknown 
$m$-tuple $(g_0,g_1,\dots,g_{m-1})$ of elements of $G$ is chosen.  
Determine the product $g_0 g_1 \cdots g_{m-1}$ of the $m$-tuple using 
queries to the oracle ${\cal O}_{(g_0,g_1,\dots,g_{m-1})}$ that acts 
on the Hilbert space $V=\C^m \otimes \C^G$ by
$$
{\cal O}_{(g_0,g_1,\dots,g_{m-1})}:|j,r\rangle\mapsto|j,g_jr \rangle. 
$$

This problem was studied for cyclic groups $G=\Z_k$ in 
[\MeyerPommersheimB].  There it was shown that the sum of $m$ elements  
of $\Z_k$ can be determined using $t$ queries with probability 
$\lfloor{m/(m-t)}\rfloor/k$.  In particular, the sum of two elements 
of $\Z_k$ can be determined with a single quantum query with 
probability $2/k$.  Our next theorem generalizes this result to an 
arbitrary group $G$.

\noindent\Theorem~2.  {\sl Let $G$ be a finite group of order $n$.  
Then there is a single query algorithm for determining the product of 
2 elements of $G$ that succeeds with probability $2/n$.}   

When $G=\Z_2$, we recover the probability 1 algorithm for Deutsch's 
problem.  In \S3, we construct the general algorithm and prove that it 
is optimal in \S4.

In \S5 we show how the solution to \SOD\ can be used to give a 
representation-theoretic interpretation of several other well-known 
quantum algorithms.  This includes the Bernstein-Vazirani algorithm 
[\BV] and van Dam's algorithm [\vanDam] for exactly identifying an 
unknown bitstring.

In \S6 we discuss why the addition of ancilla registers may improve 
the efficacy of symmetric oracle problems.

\S7 introduces a new oracle problem, \HCEP, in which we wish to 
identify a hidden group element acting by conjugation.  We see that 
(permitting ancilla registers), many groups allow identification of 
the hidden element with probability 1 in a single query.  This 
includes the dihedral and symmetric groups.

\medskip
\noindent{\bf 2.  Symmetric oracle discrimination}

In this section, we study \SOD, first proving the dimension bound and 
then proving Theorem 1, which asserts that for any representation this 
dimension bound can be achieved.  Finally, for a given representation, 
we establish exactly what choices of initial states give this optimal 
result. 

Before we begin, let us recall the following general fact, which 
limits the distinguishability of states in terms of the dimension of 
their linear span ({\it cf}.~[\MeyerPommersheim], Lemma 6.2):

\noindent\Lemma~1.  (General Dimension Bound) {\sl Suppose 
$|\psi_i\rangle$, $i=1,\ldots,n$ are $n$ pure states that span a 
space of dimension $k$.  Then any measurement to identify $i$ succeeds 
with probability at most $k/n$.}

\noindent{\sl Proof}.  Let $\rho_i$ be the density matrix 
corresponding to $|\psi_i\rangle$.  Then $\rho_i\leq \Pi_W$, 
orthogonal projection onto the span of the $n$ states.  Hence for any 
measurement $\{M_i\}$, we have
$$
{1\over n}\sum\Tr(M_i\rho_i) 
 \leq {1\over n}\sum\Tr(M_i \Pi_W) 
    = {1\over n}\Tr(\Pi_W) 
    = {k\over n}.                                      \eqno\endproof
$$

For \SOD, we have a finite group $G$ and a unitary representation 
$\Theta:G\rightarrow GL(V)$, and we are trying to distinguish among 
the group translates of an initial state $|\psi\rangle\in V$.  The 
space spanned by the orbit of $|\psi\rangle$, namely
$$
V_{\psi} = \span\{\Theta(g)|\psi\rangle\mid g\in G\},
$$
is called a {\sl cyclic subspace\/} of $V$.  Letting $d_{\Theta}$ 
denote the largest dimension of any cyclic subspace of $V$,
$$
d_{\Theta}
 = \max_{|\psi\rangle}\dim\span\{\Theta(g)|\psi\rangle\mid g\in G\},
$$
we see that Lemma 1 shows that the success probability for 
S{\eightpoint YMMETRIC} O{\eightpoint RACLE} 
D{\eightpoint ISCRIMI-} {\eightpoint NATION} is bounded above by 
$$
\psucc\leq {d_{\Theta}\over |G|}.
$$
We refer to this as the {\sl dimension bound\/} for the 
representation $\Theta$.  To understand the dimension bound better, we 
next give a formula for $d_{\Theta}$ in terms of the decomposition of 
$V$ into irreducible representations.

\noindent\Lemma~2.  {\sl Let $\Theta:G\rightarrow GL(V)$ be a finite 
dimensional representation of a finite group $G$ and let 
$$
V \cong W_1^{\oplus m_1}\oplus\cdots\oplus W_r^{\oplus m_r} 
$$
be the  decomposition of $V$ into nonisomorphic irreducible 
representations $W_k$.  Let $d_k$ be the dimension of $W_k$ and let 
$l_k=\min(m_k,d_k)$.  Then $d_{\Theta}$, the maximal dimension of a 
cyclic subspace of $V$, is given by
$$
d_{\Theta} = \sum_k l_k d_k.
$$
Furthermore, in such a cyclic subspace of $V$, the dimension of the isotypic component corresponding to $W_k$ is $d_kl_k$.
}

\noindent{\sl Proof}.  We first show that for any $|\psi\rangle\in V$, 
the cyclic subspace $V_{\psi}$ has dimension at most $\sum_k l_k d_k$.  
Since $V_{\psi}$ is generated by $|\psi\rangle$, there is a 
$\C^G$-module homomorphism $\C^G \rightarrow V_{\psi}$ that is 
surjective.  By Maschke's Theorem, $V_{\psi}$ is a summand of $\C^G$.  
Hence for each irreducible $W_k$ contained in $V_{\psi}$, the number 
$e_k$ of copies of $W_k$ satisfies $e_k\leq d_k$.  But since  
$V_{\psi}$ is a submodule of $V$, we also have $e_k\leq m_k$.  Hence, 
$e_k\leq l_k$.  It follows that 
$\dim V_{\psi}=\sum e_k d_k \leq \sum l_k d_k$, as desired. Note that equality holds if and only if $e_k = l_k$ for all $k$, that is, the dimension of the isotypic component of $V_{\psi}$ corresponding to $W_k$ is $d_kl_k$ for all $k$. This establishes the final statement.

Conversely, we must show that there exists a $|\psi\rangle\in V$ such 
that the cyclic subspace $V_{\psi}$ has dimension exactly 
$\sum l_k d_k$.  Consider the $\C^G$-module 
$W=\oplus W_k^{\oplus l_k}$.  Since $l_k\leq d_k$, there is a 
surjective $\C^G$-module homomorphism $\C^G \rightarrow W$.  Hence, 
$W$ is a cyclic $\C^G$-module.  On the other hand, since 
$l_k\leq m_k$, $W$ is isomorphic to a submodule of $V$.  Thus, $V$ has 
a cyclic submodule of dimension $\dim W=\sum l_k d_k$. \hfill\endproof

Note that for each irreducible representation $V_i$ that appears in 
$V$, up to $d_i$ copies of $V_i$ potentially contribute to the overall 
success of the algorithm; beyond that, there is no contribution.  Thus 
we may refer to $l_i= \min \{m_i, d_i\}$ as the number of ``usable 
copies'' of $V_i$.

Having established the dimension bound, we now proceed to the proof of 
Theorem~1, stated in the Introduction, which asserts that for any 
representation, the dimension bound can be achieved. 

\noindent\Definition.  An {\sl optimal input\/} for \SOD\ is a state 
$|\psi\rangle$ for which there exists a measurement distinguishing the 
symmetric states generated by $|\psi\rangle$ with optimal success 
probability, \ie,
$$
\psucc = {d_{\Theta} \over |G|}.
$$

We begin by characterizing optimal inputs, and then examine the 
characterization to prove the existence of such states.  An immediate 
simplification is to consider only representations for which 
$d_{\Theta} = \dim V$, since the dimension bound of a representation 
determines the largest possible subspace in which the measurement 
problem takes place.  Note that if $d_{\Theta} = \dim V$, then the 
analogous equation is satisfied for any subrepresentation of $V$.

\noindent\Proposition~1. {\sl Let $\Theta:G\rightarrow GL(V)$ be a 
unitary representation on the Hilbert space $V$ such that 
$d_{\Theta} = \dim V$.  Then $|\psi\rangle\in V$ is an optimal input 
if and only if
$$
\sum_{g\in G}{\Theta(g)|\psi\rangle\langle\psi|\Theta(g^{-1})} 
 = {|G|\over\dim V}I_V.                                       \eqno(1)
$$
}

\noindent{\sl Proof}.  Let $|\psi\rangle$ be an arbitrary state.  It 
is well-known that there exists an optimal measurement consisting of 
symmetric rank-1 projections distinguishing the symmetric states 
generated by $|\psi\rangle$, for instance the SRM 
[\HW,\EldarForney,\EMV].  Such a measurement is determined by a single 
measurement vector $|\mu\rangle$ satisfying the completeness relation 
$$
\sum_{g\in G}{\Theta(g)|\mu\rangle\langle\mu|\Theta(g^{-1})} = I
$$
and the success probability of the measurement is
$$
\psucc 
 = {1\over|G|}\sum_{g\in G}
   |\langle\Theta(g)\psi|\Theta(g)\mu\rangle|^2 
 = |\langle\psi|\mu\rangle|^2.
$$
By taking traces in the completeness relation, we see that 
$\|\mu\|^2 = \dim V/|G|$.  Hence by Cauchy-Schwarz,
$$
\psucc \leq \|\psi\|^2\|\mu\|^2 = {\dim V\over|G|},
$$
with equality if and only if $\psi = \lambda\mu$ for some scalar 
$\lambda\in\C$.

If this is true, then the completeness relation implies (1).  
Conversely, if (1) is satisfied then the vector 
$|\mu\rangle = \sqrt{\dim V/|G|}|\psi\rangle$ describes a valid 
symmetric measurement and the success probability is $\dim V/|G|$.                                
                                                       \hfill\endproof

An interesting corollary to Proposition 1 is \SOD\ in the case of an 
irreducible representation.

\noindent\Corollary~1.  {\sl Let $\Theta:G\rightarrow GL(V)$ be an 
irreducible representation.  Then any choice of initial state is an 
optimal input.}

\noindent{\sl Proof}.  For any $|\psi\rangle\in V$, the operator 
$\sum_{g\in G}{\Theta(g)|\psi\rangle\langle\psi|\Theta(g^{-1})}$ 
commutes with each $\Theta(h)$; hence it is a $\C^G$-module 
homomorphism.  By Schur's Lemma, it equals a scalar multiple of the 
identity, and taking traces yields the correct scalar.  By 
Proposition~1, $|\psi\rangle$ is optimal for \SOD.     \hfill\endproof

Next we examine the characterization (1). In the following, the symbol $T_{\psi}$ denotes the operator $\sum_{g \in G}{\Theta(g)|\psi \rangle \langle \psi |\Theta(g)^{-1}}$ for a given $|\psi \rangle$. Note that $T_{\psi} = {|G| \over \dim V}I_V$ if and only if $T_{\psi} = \lambda I_V$ for some scalar $\lambda \in \C$, since the usual trace argument deduces $\lambda = {|G| \over \dim V}$, so in checking (1) it suffices to show $T_{\psi}$ is a scalar multiple of $I_V$.

We are now ready to prove Theorem 1, which asserts that for any representation, there exists an optimal input for \SOD. We proceed in two steps, first reducing to the case that the representation consists of a single isotype of irreducible subrepresentations, and then providing a characterization of optimal inputs in such representations that clearly describes the construction of such states.

\noindent\Proposition\  2.
Let  $\Theta:G\rightarrow GL(V)$ be a unitary representation on the Hilbert space $V$  and let 
$$
V= V_1\oplus \cdots \oplus V_r \cong m_1W_1  \oplus \cdots \oplus m_rW_r 
$$
be the orthogonal decomposition of $V$ into its isotypic components $V_i \cong m_iW_i$ (where $W_i$ is irreducible). Let $d_k=\dim W_k$, and assume that $m_k\leq d_k$ for all $k=1,\dots,r$. Suppose $|\psi \rangle$ is an arbitrary state and write $|\psi \rangle = |\psi_1 \rangle + \dots + |\psi_r \rangle$ with respect to the canonical decomposition above. 

Then $|\psi \rangle$ is an optimal input if and only if $\|\psi_i\|^2 = {\dim V_i \over \dim V}$ and the states ${\sqrt{\dim V \over \dim V_i}}|\psi_i \rangle$ are optimal inputs with respect to the subrepresentations $V_i$.

\medskip

\noindent{\bf Proof.} First suppose $|\psi \rangle$ satisfies (1). This equation is invariant under the orthogonal projection $\Pi_i$ corresponding to $V_i$. That is,
$$
\sum_{g \in G}{\Theta(g)|\Pi_i\psi \rangle \langle \Pi_i\psi |\Theta(g^{-1})} = {|G| \over \dim V}I_{V_i}
$$
Taking traces we see $\|\psi_i\|^2 = {\dim V_i \over \dim V}$, and normalizing the above equation shows that each state ${\sqrt{V \over \dim V_i}}|\psi_i \rangle$ satisfies (1) with respect to the representation $V_i$.

Conversely, consider $|\psi \rangle = |\psi_1 \rangle + \dots + |\psi_r \rangle$ and compute
$$
T_{\psi} = \sum_{g \in G}{\sum_{i,j = 1}^r{\Theta(g)|\psi_i \rangle \langle \psi_j |\Theta(g^{-1})}}
$$
Since $\sum_{g \in G}{\Theta(g)|\psi_i \rangle \langle \psi_j|\Theta(g^{-1})}$ is a $\C^G$-module homomorphism between $V_j$ and $V_i$, the above reduces to
$$
T_{\psi} = \sum_{i = 1}^r\sum_{g \in G}{\Theta(g)|\psi_i \rangle \langle \psi_i|\Theta(g^{-1})}
$$
So if each summand here is equal to ${|G| \over \dim V}I_{V_i}$ (as the conditions imply), we have that $T_{\psi} = {|G| \over \dim V}I_V$, which is exactly condition (1).
\hfill \endproof

Proposition 2 reduces the question of the existence of optimal inputs for general representations to that of representations with single isotypes of irreducible representations. Thus we tackle this case in the next Proposition to prove Theorem 1.

\noindent\Proposition\  3.
Let  $\Theta:G\rightarrow GL(V)$ be a representation on the Hilbert space $V$ such that $V \cong W^{\oplus m}$ for some irreducible representation $W$.  Let $d=\dim W$ and assume that $m\leq d$.   Denote by $\alpha_i:W\rightarrow V$ the inclusion of $W$ into the $i$th coordinate of $V$. Then a state $|\psi \rangle \in V$ is an optimal input if and only if $|\psi\rangle$ is of the form
$$
|\psi\rangle = {1 \over {\sqrt{m}}} \sum_{i=1}^m \alpha_i(|e_i\rangle), 
$$
where $\{|e_1\rangle,\cdots,|e_m\rangle\}$ is an orthonormal set in $W$.

\medskip

\medskip
\noindent{\bf Proof.}
Let $\pi_i:V\rightarrow W$ be the $i$th projection map. We will denote by $\rho$ the representation of $G$ on $W$. As we used in Corollary 1, $T_{\psi}$ is a $\C^G$-module endomorphism of $V$. Taking a basis of $V$ formed from successive bases of the irreducible components and applying Schur's Lemma, we can write $T_{\psi}$ as a block matrix whose $(i,j)$th component matrix is a scalar times the identity. Let $|e_i\rangle = \pi_i(|\psi\rangle)$ so $|\psi\rangle = \sum_{i=1}^m{\alpha_i(|e_i\rangle)}$. Then noting that the $(i,j)$th component of $T_{\psi}$ is $\pi_iT_{\psi}\alpha_j$, we compute
$$
\pi_i T_{\psi} \alpha_j = \pi_i \sum_{g \in G}\sum_{s,t=1}^m{ \Theta(g) |\alpha_s e_s\rangle \langle \alpha_t e_t| \Theta(g)^{-1} \alpha_j}
$$
Since $\pi_i$ and $\alpha_j$ are $\C G$-module homomorphisms, this becomes
$$
\pi_i T_{\psi} \alpha_j = \sum_{g \in G}{\rho_g |e_i \rangle \langle e_j| \rho_g^{-1}}
$$
where we used that $\pi_i\alpha_s = \delta_{i,s}I_W$ and $\langle \alpha_t e_t|\alpha_j = \delta_{t,j}\langle e_j |$. Now this component must be a scalar multiple of $I_W$, and taking traces we see that the scalar is ${|G|\langle e_i | e_j \rangle \over d}$.

Now $T_{\psi}$ is a scalar multiple of the identity (with scalar $\lambda$) if and only if $\pi_i T_{\psi} \alpha_j = \delta_{i,j}\lambda I_W$ which occurs if and only if $\langle e_i | e_j \rangle = \delta_{i,j}{d\lambda \over |G|}$. This condition means exactly that $\{e_1, \dots e_m\}$ form an orthogonal set of vectors in $V$ of length $\sqrt{d\lambda \over |G|}$. Now if the equation $T_{\psi} = \lambda I_V$ holds then taking traces we see $\lambda = {|G| \over md}$. Hence rescaling the vectors $\{e_1, \dots e_m\}$ gives us the needed decomposition of $|\psi\rangle$.
\hfill \endproof

This concludes the proof of Theorem 1, since we may construct an optimal input from an orthogonal invariant decomposition of the representation and a choice of orthonormal states in each irreducible isotype: in each canonical summand an optimal state is constructed as in Proposition 3, and scalar weights are assigned to each of these according to Proposition 2 to produce the final state. For optimality, the weights must have norm squared proportional to the dimension of the summand, but otherwise the relative phases are irrelevant. In fact, we have characterized optimal inputs as those states arising from all such constructions. The construction is used explicitly in Lemma 3 below.


\medskip \break
\noindent{\bf 3.  Quantum Group Multiplication}

In this section we prove Theorem 2, giving a single-query quantum algorithm for multiplying two elements of a group $G$ with probability ${2 \over |G|}$.

Before beginning the proof of Theorem 2, we make an observation about our solution to the \SOD\ problem given in the previous section.  For that problem, we have constructed an optimal initial state and measurement, and we know the probability of measuring the correct element $g\in G$.   It turns out that it is also useful to know the probability of measuring some other element $h$ given the the correct element is $g$.  The following lemma gives a formula of this probability.

\medskip
\noindent\Lemma\ 3.  Let $G$ be a finite group of order $n$ and let $V$ be a representation of $G$.  Assume that  when $V$ is written as a direct sum of irreducible representations, each irreducible $V_k$ appearing in the decomposition appears exactly $d_k = \dim V_k$ times. (We leave open the possibility that some irreps do not appear at all in the decomposition.)  Let $|\psi\rangle$ be an optimal input, with $\mu = \sqrt{\dim V \over |G|}|\psi \rangle$ describing the optimal symmetric measurement.  Then $P(h|g)$, the probability of measuring $h$ when the chosen element is actually $g$ is given by
$$
P(h|g) = {1 \over {\dim V |G|}} | \chi_V(h^{-1}g) |^2= {1 \over {\dim V |G|}} \biggl| \sum_k d_k \chi_k(h^{-1}g) \biggr|^2,
$$
where $\chi_k$ is the irreducible character corresponding to $V_k$.

\medskip
\noindent{\bf Proof.}\   First note that if $\Theta$ denotes the given representation of $G$ on $V$, then it follows that $d_{\Theta}=\dim V=\sum d_k^2.$  Recall the measurements vectors determined by $\mu$ are $\mu_h = \Theta(h)\mu$. Hence
$$
P(h|g)= |\langle \mu_h|\psi_g\rangle |^2 =  {d_{\Theta} \over { |G|}} |\langle \psi_h|\psi_g\rangle |^2 = {d_{\Theta} \over { |G|}} |\langle \psi|\Theta(h^{-1}g)|\psi\rangle |^2.
$$
To examine this more carefully, we describe $|\psi \rangle$ according to the construction outlined in Propositions 2 and 3. Fix an orthogonal decomposition of $V$ into irreducibles with sections $\alpha_{k,i}$ including the $i$th copy of $V_k$ into $V$ and similarly proejctions $\pi_{k,i}: V \to V_k$. By Propositions 2 and 3, in each $V_k$ the vectors $\{e_{k, i} := \sqrt{d_{\Theta} \over d_k}\pi_{k,i}(\psi)\}_{i=1}^{d_k}$ form an orthonormal basis. We denote $f_{k,i} = \alpha_{k,i}(e_{k,i})$ so that $|\psi \rangle = \sum_{k,i}{\sqrt{d_k \over d_{\Theta}}f_{k,i}}$. Finally let $q = h^{-1}g$. Then we obtain
$$
P(h|g) = {1 \over {{d_{\Theta}} |G|}} \biggl| \sum_{k,i,k',i'} \sqrt{d_k d_{k'} } \langle f_{k',i'} | \Theta(q) | f_{k,i} \rangle      \biggr|^2.
$$
Since $f_{k,i}$ lives in the $i$th copy of $V_k$ in $V$, which is an invariant subspace, every term in the above sum vanishes unless $k=k'$ and $i=i'$.  Thus we obtain
$$
P(h|g) = {1 \over {{d_{\Theta}} |G|}} \biggl| \sum_{k,i} d_k  \langle f_{k,i} | \Theta(q) | f_{k,i} \rangle      \biggr|^2.
$$
Letting $\rho_k:G\rightarrow GL(V_k)$ be the irreducible representation of $G$ on $V_k$, we see that $\alpha_{k,i}$ commutes with the action of $G$; that is, $\Theta(g) \circ \alpha_{k,i} = \alpha_{k,i} \circ \rho_k(g)$.   Hence the above inner products may be computed in  $V_k$ as
 $$
P(h|g) = {1 \over {{d_{\Theta}} |G|}} \biggl| \sum_{k,i} d_k  \langle e_{k,i} | \rho_k(q) | e_{k,i} \rangle      \biggr|^2.
$$              
For each $k$, $\{e_{k,1}, \dots, e_{k,d_k}\}$ is an orthonormal basis of $V_k$.  Hence the above sum computes the trace $\chi_k(q)$ of $\rho_k(q)$.  That is, we obtain
$$
 P(h|g) = {1 \over {{d_{\Theta}} |G|}} \biggl|   \sum_k d_k \chi_k(q)   \biggr|^2. $$ \hfill \endproof

We now prove Theorem 2.  We wish to determine the product of two elements $g_0,g_1\in G$  assuming  access to an oracle $\calo_{g_0,g_1}$, which acts on $W =\C^2\otimes\C^G$ by
$$
\calo_{g_0,g_1} : |i,s\rangle \mapsto |i, g_i s\rangle.
$$
This defines an representation of the group $G\times G$ on $W$.   The representation decomposes into two invariant subspaces:
$$ W = |0\rangle \otimes \C^G \oplus |1\rangle \otimes \C^G,$$ 
and each of these two subspaces decomposes like the regular representation of $G$, with $|0\rangle \otimes \C^G$ containing  $d_i$ copies of the irreducible representation $V_i$ where the left factor of $G\times G$ is acting, and $|1\rangle \otimes \C^G$ containing  $d_i$ copies of the irreducible representation $V^\prime_i$ where the right factor of $G\times G$ is acting.  Note that there are two copies of the trivial representation of $G\times G$, but otherwise the $V_i$ and $V^\prime_i$ are non-isomorphic as representations of $G\times G$.

Throw away the two copies of the trivial representation, and call the remaining space $V$.  We then have $\dim V = 2(n - 1)$, and $V$ decomposes as $d_i$ copies of the $V_i$ and $d_i$ copies of $V^\prime_i$.  We apply Lemma 3 to this situation to find:
$$
P((h_0,h_1)|(g_0,g_1)) = {1 \over {2(n-1) n^2}} \biggl| \sum_i d_i \chi_i (h_0^{-1} g_0) + \sum_i d_i \chi_i (h_1^{-1} g_1) \biggr|^2,
$$
where the sums are taken over all nontrivial irreducible characters of $G$.   Using column orthogonality, this simplifies to 
$$
P((h_0,h_1)|(g_0,g_1)) = {1 \over {2(n-1) n^2}} (tn-2)^2,
$$
where $t\in\{0,1,2\}$ counts the number of $j$ such that $h_j = g_j$.

When $t=2$, we are guaranteed a correct product; when $t=1$, we are guaranteed an incorrect product; when $t=0$, we have a $1 \over {n-1}$ chance of a correct product.  There are $(n-1)^2$ pairs $(h_0, h_1)$ for which $t=0$.  Thus the total success probability of getting the right product is   
$$
P_{succ} = {1 \over {2(n-1) n^2}} (2n-2)^2 + {1 \over {2(n-1) n^2}} (-2)^2 {1 \over{ n-1}}  (n-1)^2 = {2\over n},
$$
as desired.
\hfill\endproof

\medskip
 {\bf Any single nontrivial irreducible representation works for group multiplication.} \  The above single-query algorithm for group multiplication uses all the non-trivial representations of $G\times G$ appearing in $\C^2\otimes\C^G$.  We now show that the same success probability can be achieved by picking any nontrivial irreducible representation $W$ of $G$ and choosing the optimal query in the subspace $|0\rangle\otimes W^{\oplus d} \oplus |1\rangle \otimes (W^*)^{\oplus d}\subset \C^2\otimes\C^G$, where $d=\dim W$ and  $W^*$ is the dual representation.

\noindent\Theorem\  4.
Let $G$ be a group of order $n$.  Let $W$ be any nontrivial irreducible representation of $G$, and let $d=\dim W$.  Let $Y\subset \C^G$ be the $W$-isotypic subspace of $\C^G$, i.e., the unique $G$-invariant subspace with $Y\cong W^{\oplus d}$. Let $W^*$ denote the dual representation of $W$, and let $Y^*$ denote the $W^*$-isotypic subspace of $\C^G$.   Let $V = |0\rangle\otimes Y \oplus |1\rangle\otimes Y^* \subset \C^2\otimes\C^G$.   Consider the \SOD\ problem for  the group $\Gamma= G\times G$ acting on $V$, and use an initial state and measurement as described in Propositions 1, 2 and 3.  Then this measurement returns a pair of elements of $G$, and the product of these two elements is correct with probability $2/n$.

\medskip
\noindent{\bf Proof.} 
We use Lemma 3 to compute the probability $P((h_0,h_1)|(g_0,g_1))$ of measuring a given pair $(h_0,h_1)$ given that the hidden pair is $(g_0,g_1)$.   Letting $q=(q_0,q_1)=(h_0^{-1}g_0, h_1^{-1}g_1)$, and noting that $\dim V=2d^2$, we find
$$
P((h_0,h_1)|(g_0,g_1)) = { 1 \over 2d^2n^2 } |d\chi_W(q_0) + d\chi_{W^*}(q_1)|^2 = { 1 \over 2n^2 } |\chi_W(q_0) + \overline{\chi_{W}(q_1)}|^2.
$$
For fixed $(g_0, g_1)$, we wish to sum the above quantity over all pairs $(h_0,h_1)$ such that $h_0h_1=g_0g_1$.   This condition is satisfied if and only if $q_1=g_1^{-1}q_0^{-1}g_1$.  For a fixed $q_0$, there is a unique $q_1$ with this property.  Thus we see that the success probability for measuring a pair $(h_0,h_1)$ with the correct product is
$$
P_{\rm{correct\ product}} = { 1 \over 2n^2 }  \sum_{q_0\in G}  |\chi_W(q_0) + \overline{\chi_{W}(g_1^{-1}q_0^{-1}g_1)}|^2.
$$
Since the value of a character is invariant under conjugation and $\chi_W(q_0^{-1})=\overline{\chi_W(q_0)}$, we obtain
$$
P_{\rm{correct\ product}} = { 1 \over 2n^2 }  \sum_{q_0\in G}  |2\chi_W(q_0) |^2={ 2 \over n^2} \sum_{q_0\in G} |\chi_W(q_0) |^2 = {2 \over n}. \hskip 10em \endproof$$

\bigskip
\noindent{\bf 4.  Optimality of single-query group multiplication algorithm.}

We now prove the optimality of the single-query group multiplication algorithm given in the previous section. To do this, we rephrase \GM\ as a mixed state discrimination problem where each mixed state corresponds to a possible hidden product. For a given initial state $| \psi \rangle$, we consider the mixed states:
$$
\rho_g = {1 \over |G|}\sum_{g_0g_1=g}{\calo_{g_0,g_1}| \psi \rangle \langle \psi | \calo_{g_0,g_1}^{-1}}
$$
The task of designing an efficient algorithm for \GM\ is equivalent to picking an initial state $| \psi \rangle$ and a $G$-valued POVM that distinguishes these mixed states. Therefore, optimality of the algorithm above is provided by:

\noindent\Theorem\ 5. For any inital state $| \psi \rangle$, any measurement distinguishing the mixed states $\{\ \rho_g\ |\ g \in G\}$ succeeds with probability at most ${2 \over |G|}$.  

\noindent{\bf Proof.} Fix an initial state $| \psi \rangle$. Note that the mixed states enjoy the following symmetry, which is geometric uniformity for mixed states:
$$
\rho_g = \calo_{g,e}\rho_e\calo_{g,e}^{-1}
$$
where $e \in G$ is the identity element. Eldar et al. [\EMV]\ show that there exists an optimal measurement which also exhibits this symmetry. Hence we consider a POVM $\{X_g\ |\ g \in G\}$ satisfying
$$
X_g = \calo_{g,e}X_e\calo_{g,e}^{-1}
$$
This symmetry combined with the completeness relation imposes some restrictions on the shape of the measurement operators. Using the decomposition of our space as $|0\rangle \otimes \C^G \oplus |1\rangle \otimes \C^G$ and picking the canonical basis in each summand, we may write $X_e$ as a $2 \times 2$ block matrix
$$
X_e = \left( \matrix{ A & C \cr C^* & B} \right)
$$
where $A, B, C$ are $|G| \times |G|$ matrices and $A, B$ are positive. Letting $L_g$ denote the matrix of left multiplication by $g$ in $\C^G$, we have
$$
X_g = \calo_{g,e}X_e\calo_{g,e}^{-1} 
= \left( \matrix{ L_gAL_g^* & L_gC \cr C^*L_g^* & B } \right)
$$
Hence the completeness relation implies
$$
\sum_{g \in G}{L_gAL_g^*} = I \quad \mbox{and} \quad B = {I \over |G|}
$$
We are ready to examine the success probability.
$$
\psucc = {1 \over |G|}\sum_{g \in G}{\Tr (\rho_gX_g)} 
= {1 \over |G|}\sum_{g \in G}{\calo_{g,e}\rho_e\calo_{g,e}^*\calo_{g,e}X_e\calo_{g,e}^*} = \Tr(\rho_eX_e)
$$
Now expanding the sum in $\rho_e$, we find
$$
\psucc = {1 \over |G|}\Tr \left(\sum_{h \in G}{\calo_{h,h^{-1}}^*|\psi \rangle \langle \psi |\calo_{h,h^{-1}}X_e} \right)
= {1 \over |G|}\langle \psi | \left( \sum_{h \in G}{\calo_{h,h^{-1}}X_e\calo_{h,h^{-1}}^*} \right) | \psi \rangle
$$
Let $X = \sum_{h \in G}{\calo_{h,h^{-1}}X_e\calo_{h,h^{-1}}^*}$. It is a positive operator. Then
$$ \psucc = {1 \over |G|}\langle \psi | X \psi \rangle $$
We can estimate this quantity by examining the structure of $X$ as a $2 \times 2$ block matrix. Using our previous notation for $X_e$, we have
$$
\calo_{h,h^{-1}}X_e\calo_{h,h^{-1}}^* =
 \left( \matrix{L_hAL_h^* & L_hCL_{h^{-1}}^* \cr
  L_{h^{-1}}C^*L_h^* & L_{h^{-1}}BL_{h^{-1}}^*} \right)
$$
Now our knowledge about $A$ and $B$ from the completeness relation gives
$$
X = \left( \matrix{I & D \cr D^* & I} \right)
$$
where $D = \sum_{h \in G}{L_hCL_{h^{-1}}^*}$. A lemma from basic matrix analysis ([\Bhatia] Proposition 1.3.1) states that $X$ is a positive operator if and only if $||D|| \leq 1$, where $||\cdot ||$ denotes the operator norm. With this in hand we perform the final estimation of the success probability. Let $| \psi \rangle = | 0, \psi_0 \rangle + |1, \psi_1 \rangle$ be the decomposition of $| \psi \rangle$ into its components in each direct summand. Then we may calculate directly:
$$
\eqalign{\langle \psi | X \psi \rangle &= \langle \psi_0 | \psi_0 \rangle + \langle \psi_1 | \psi_1 \rangle + \langle \psi_0 | D \psi_1 \rangle + \langle \psi_1 | D^*\psi_0 \rangle \cr
&= 1 + 2\Re \langle \psi_0 | D \psi_1 \rangle \cr
&\leq 1 + 2|\langle \psi_0 | D \psi_1 \rangle| \cr
&\leq 1 + 2\|\psi_0\|\|D\psi_1\|\cr
&\leq 1 + 1}
$$
Therefore $\psucc = {1 \over |G|}\langle \psi | X \psi \rangle \leq {2 \over |G|}$. \hfill \endproof

The above inequalities invite a partial converse to Theorem 4, which provided a solution to \GM\ using a pair of nontrivial dual representations in the two copies of $\C^G$. Let $W_1, \dots, W_r$ denote the canonical summands of $\C^G$ with projections $\pi_i: V \to W_i$. Then we have

\noindent\Proposition\  4. Suppose $|\psi \rangle$ is an input state that solves \GM\ with probability ${2 \over |G|}$. Write $|\psi \rangle = \psi_0 + \psi_1$ where each $\psi_i \in \C^G$. Then the following holds for each canonical summand $W_i$:
$$
\|\pi_i(\psi_0)\| = \|\pi_i^*(\psi_1)\|
$$
where $\pi_i^*: V \to W_i^*$ is the canonical projection to the summand whose isotype is dual to the isotype of $W_i$.

\noindent{\bf Proof.} The statement is a consequence of the conditions for equality in each of the inequalities used in the previous proof. Combined, they state that $P_{\rm success} = {2 \over |G|}$ if and only if $\psi_0 = D\psi_1$.

We examine $D$ more properly: recall it is defined as
$$
D = \sum_{h \in G}{hCh}
$$
where for our purposes now, $C$ is an arbitrary linear operator and $h$ denotes its action as a linear operator on $\C^G$. As opposed to the operator $T$ used in Proposition 3, $D$ is not a $\C^G$ module homomorphism. However, let $e_i$ denote the orthogonal idempotent of $\C^G$ corresponding to $W_i$. It is well-known that
$$
e_i = {1 \over |G|}\sum_{g \in G}{\chi_i(1)\chi_i(g^{-1})g}
$$
where $\chi_i$ is the character corresponding to the irreducible isotype of $W_i$. Then a simple computation yields:
$$
e_iD = D{e_i}^*
$$
where ${e_i}^* = {1 \over |G|}\sum_{g \in G}{\chi_i(1)\chi_i(g)g}$ is the idempotent corresponding to $W_i^*$. Therefore $D$ leaves invariant the vector space direct sum decomposition $\C^G = W_1 \oplus \dots \oplus W_r$, taking a summand to its dual. Therefore if $\psi_0 = D\psi_1$, then the part of $\psi_0$ in $W_i$ is the same length as the part of $\psi_1$ in $W_i^*$. This concludes the proof.\hfill \endproof

\bigskip
\noindent{\bf 5.  Relation to Abelian Quantum Learning Problems}

In this section we recall some well-known abelian quantum learning problems and discuss how the results concerning \SOD\ provide a representation theoretic explanation of the construction and optimality of these algorithms, in particular the Bernstein-Vazirani and van Dam algorithms.

{\bf The Bernstein-Vazirani problem.}\
The Bernstein-Vazirani search problem is the task of identifying an element $a \in \Z_2^n$ given an oracle that returns $a \cdot x \ \mod 2$ when queried by $x \in \Z_2^n$ [\BV]. These oracles are highly symmetric; so much so in fact, that there is an algorithm that uses only a single query to identify the hidden string with certainty. The algorithm is summarized as follows: the oracles $\calbv_a$ act on $(\C^2)^n \otimes \C^2$ by
$$
\calbv_a|x, i \rangle = |x, i \oplus (a \cdot x) \rangle
$$
Let $H^{\otimes n}$ denote the $2n \times 2n$ Hadamard matrix and $\eta_n = H^{\otimes n}|0 \rangle$ the equal superposition vector. A phase kickback calculation shows that by applying the oracle to the state $|\eta_n, -\rangle$ followed by a Hadamard transform on the query register, we may measure this register in the computational basis to witness the outcome $a$ with certainty. This is expressed by the following equation:
$$
|a, - \rangle = \left( H^{\otimes n} \otimes I \right)\calbv_a|\eta_n, - \rangle
$$
This problem is directly addressed by our solution to \SOD\ once we notice that the map $a \mapsto \calbv_a$ is a unitary representation of $\Z_2^n$ with the canonical decomposition
$$
(\C^2)^n \otimes \C^2 = \bigoplus_{a \in \sZ_2^n \atop a \neq 0}|a, -\rangle \oplus W_0
$$
where $W_0$ is an invariant subspace consisting of $2^n + 1$ copies of the trivial representation. The remaining subrepresentations comprise a full set of non-trivial isomorphism types of irreducible representations of $\Z_2^n$. Therefore, the dimension bound tells us that the oracles can be distinguished with probability $1$ in a single query.

Furthermore, in this scenario the initial state constructed in Section 2 consists of equal contributions from each isomorphism type of irreducible representations. For the non-trivial representations, there is only one choice (up to phase), but for the trivial representation, any vector of suitable length taken from $W_0$ may be chosen. In the above (and typical) presentation of the algorithm, the contribution from the trivial subspace is taken to be $|0, - \rangle$. However, we note that this choice, as well as the relative phases of these contributions may differ from the equal superposition, and the algorithm will still yield a perfect result.

{\bf The van Dam algorithm.} The van Dam problem also considers the task of learning an element of $\Z_2^n$ but uses much less sophisticated oracles and makes use of multiple (non-adaptive) queries [\vanDam]. We consider a hidden element as a function $f: \{0, \dots ,n-1\} \to \{0, 1\}$ and the oracles encode ``evaluation queries'', i.e. a single query consists of an element of the domain $i \in \{0, \dots, n-1\}$ and the response is $f(i)$. Van Dam's algorithm requires multiple queries to identify the function within a small probability of error and consists of two steps; the first step approximates the Bernstein-Vazirani oracles, and the second step may be interpreted as an instance of \SOD\ and thus we may apply our framework to this stage.

The first step consists of using $k$ queries to construct a new oracle $A_k(f)$ acting on $(\C^2)^n \otimes \C$ via:
$$
A_k(f)|x, b \rangle = \cases{|x, b \oplus (x \cdot f) \rangle &if $\|x\| \leq k$ \cr
|x, b \rangle &if $\|x\| > k$\cr}
$$
Here $\|x\|$ denotes the Hamming weight of $x \in \Z_2^n$. It is easy to check that $f \mapsto A_k(f)$ is a unitary representation of $\Z_2^n$, and the case $k = n$ reduces to the Bernstein-Vazirani oracles.

The second stage of the algorithm consists of using the new oracles $A_k(f)$ in a single query paradigm to guess the element $f$. Van Dam continues by choosing as an input state $|\eta_k, -\rangle$, where $|\eta_k \rangle$ is the equal superposition of those standard basis elements with Hamming weight less than or equal to $k$. Upon querying the oracle $A_k(f)$, an $n$-fold Hadamard transform $H^{\otimes n}$ is applied to the first $n$ qubits and then this register is measured in the computational basis to yield the hidden element $f$ with probability
$$
\psucc = {1 \over 2^n}\sum_{i=0}^k{n \choose i}
$$

We now analyse the second stage of this algorithm and show that given the oracles $A_k(f)$, the provided algorithm is optimal. First we calculate the dimension bound. To that end, we see that the canonical decomposition of the representation is given by
$$
(\C^2)^n \otimes \C^2 = \bigoplus_{{x \in \sZ_2^n \atop \|x\| \leq k} \atop x \neq 0}{|x, -\rangle} \oplus W_0
$$
where $W_0$ is a trivial representation of dimension $2^{n+1} - \sum_{i=1}^k{n \choose i}$. Since $\Z_2^n$ is abelian, each isomorphism type of irreducible representation contributes only $1$ to the dimension bound, so we have a maximal success probability of ${1 \over 2^n}(\sum_{i=1}^k{n \choose i} + 1) = {1 \over 2^n}\sum_{i=0}^k{n \choose i}$ which is the success probability of van Dam's algorithm.

Concerning the input state, the same comments may be made as in the Bernstein-Vazirani problem: the input state chosen by van Dam is an instance of the initial state constructed in Section 2, but apparently there is some flexibility in choosing a contribution from $W_0$ and the relative phases of all contributions that lead to a similar result.

These examples suggest a general strategy for constructing multi-query algorithms, which is to spend a certain number of queries setting up a symmetric single-query problem and then apply the solution to \SOD.

\bigskip
\noindent{\bf 6.  Symmetric Oracle Identification with Ancilla Register}

In many situations in quantum computing, one is allowed to use an ancilla register.  Consider the \SOD\ problem for some representation $V$ of a group $G$. If we introduce an $r$-dimensional ancilla register $Z=\C^r$, our new Hilbert space is $V\otimes Z$ and the oracle acts by $\Theta(g)\otimes I_Z.$  

Tensoring with an $r$-dimensional trivial representation has the effect of multiplying by $r$ the number of copies of each irreducible representation contained in $V$. That is, if $W_k$ is an irreducible representation and $m_k$ is the number of copies of $W_k$ contained in $V$, then $V\otimes Z$ contains $rm_k$ copies of $W_k$.  We thus have the following proposition, which follows directly from Theorem 1 and Lemma 2.

\noindent\Proposition\  5. Let   $\Theta:G\rightarrow GL(V)$  be a finite dimensional representation of a finite group $G$.  and let 
$$
V\cong W_1^{\oplus m_1} \oplus W_2^{\oplus m_2} \oplus \cdots \oplus W_r^{\oplus m_r} 
$$
be the  decomposition of $V$ into  nonisomorphic irreducible representations $W_k$.  Let $d_k$ be the dimension of $W_k$.
Then using an $r$-dimensional ancilla register, the optimal success probability for \SOD\ is given by 
 $$
 P_{\rm{success}}= {1 \over |G|}\sum \min(rm_k,d_k)d_k. 
 $$
Hence, if the size of the ancilla register is large enough, one can succeed with probability 
$$
 P_{\rm{success}}=  {1 \over |G|}\sum  d_k^2,
 $$
where the sum is taken over all the irreducible representations that appear in $V$ with positive multiplicity.

\noindent\Corollary\  2. Let $V$ be a representation of $G$ and consider the \SOD\ problem for $V$ allowing as many ancilla qubits as desired.  Then there exists a single-query probability $1$ solution to this problem if and only if every irreducible representation of $G$ appears in $V$ with positive multiplicity.

\bigskip
\noindent{\bf 7.  Hidden Conjugating Element Problem}

In this section we consider the \SOD\ problem for the action of a group on itself by conjugation.   If $G$ is any finite group, then $G$ acts on $V=\C^G$ by
$$
\Theta(g)(|h\rangle)=|ghg^{-1}\rangle.
$$
This defines a representation of $G$ and we consider the \SOD\ problem for this representation. That is, we wish to determine a hidden element $g\in G$ and we are given access to an oracle that a query $x\in G$ with $gxg^{-1}$.  We refer to this problem as the \HCEP.

Classically, with a single query $x$, we may determine $g$ with probability   
$$
 P_{\rm{classical}} = {|cl(x)|\over |G|},
$$
where $cl(x)$ denotes the conjugacy class of $x$.  Thus, the optimal query is an element $x\in G$ with the largest number of conjugates. 

We also observe that if $g$ belongs to the center $Z(G)$ of the group, i.e., $g$ commutes with every element of $g$, then $\Theta(g)=I_V$.   Thus if $Z(G)\neq\{e\}$, then it is impossible to determine $g$ no matter how many classical or quantum queries are used.  Thus we will assume that $Z(G)=\{e\}$.   (Alternatively one could work in the group $G/Z(G)$.)

As we have seen, we can determine the optimal success probability for this problem if we know how to decompose the conjugation representation into irreducible representations.  

\noindent\Lemma\ 4.  Let $G$ be a finite group and let $V$ be the conjugation representation defined above. Let $W$ be any irreducible representation with character $\chi_W$ and  $m_W$  be the multiplicity of $W$ in $V$.  Then $m_W$ equals the sum of the $\chi_W$ row of the character table of $G$.  That is,
$$
m_W= \sum \chi_W(g),
$$ 
where the sum is computed by choosing one element from each conjugacy class of $G$.  

\noindent{\bf Proof.}  Since $V$ is a permutation representation, the character value $\chi_V(g)$ is the number of fixed points of $g$, i.e.,
$$
\chi_V(g) = |C_G(g)|={  |G|  \over |cl(g)|},
$$
where $C_G(g)$ is the centralizer of $g$, i.e., all elements of $G$ that commute with $g$.  Hence,
$$
m_W=\langle \chi_V, \chi_W \rangle = {1 \over |G|} \sum  |cl(g)| {  |G|  \over |cl(g)|} \chi_W(g) = \sum \chi_W(g),
$$ 
where all sums are computed using one element from each conjugacy class of $G$.
\hfill \endproof

We now consider the dihedral group $D_n$ of order $2n$.   Here we assume that $n$ is odd, since if $n$ is even, $Z(D_n)$ is nontrivial.   For odd $n$ we show that \HCEP\ for $D_n$ has a single query probability one solution.

\noindent\Theorem\ 7.  If $G=D_n$ is the dihedral group of order $n$, where $n$ is a positive odd integer.    Then using a single ancilla qubit, there is a probability one algorithm for \HCEP.

\noindent{\bf Proof.}  
From the character table of $D_n$ and Lemma 4, we find that the conjugation representation $V$ contains ${n+3\over 2}$ copies of the trivial representation, ${n-1\over 2}$ copies of the one-dimensional alternating representation, and a single copy of each of the ${n-1}\over 2$ two-dimensional irreducible representations.  After tensoring with two-dimensional ancilla register, we find that every irreducible representation $W$ of $D_n$ appears at least $\dim W$ times.  It follows that there is a single query algorithm that succeeds with probability one.
\hfill \endproof

Note that classically, the best we can do with one query is probability $p=1/2$.  Also note that  without an ancilla qubit, we find $d_{\Theta}={n+1\over 2n}$, so the best we can do with a single quantum query is $p=1/2+1/2n$, only slightly better than classical.

For the symmetric group $S_n$, we also obtain a probability one single-query algorithm for any $n$.  In contrast, the optimal classical algorithm succeeds with probability $1/n$.

\noindent\Theorem\  8.
 Consider \HCEP\ for the symmetric group $S_n$, allowing ancilla qubits. For any $n$, there is a single-query probability $1$ algorithm.

\noindent{\bf Proof.}  Frumkin [\Frumkin] showed that the conjugation representation for $S_n$ contains at least one copy of every irreducible representation of $S_n$.  Hence the theorem follows from Proposition 5.
\hfill \endproof

 As we have seen, for certain classes of groups, the conjugation representation contains every irreducible representation.  The question of which groups have this property was investigated in 1971 by Roth [\Roth], who gave families of groups for which this is true. Roth further conjectured that this is case for any group, provided that the center is trivial.  Shortly after Roth's conjecture, however, Formanek [\Formanek] found a counterexample.  Hence, there are groups $G$ with trivial center for which  \HCEP\ cannot be solved with a single query, even allowing as many ancilla qubits as desired. 
 
 In spite of this, it still seems possible that the representations left out of the conjugation representation are scarce.  In this case, by Proposition 5, the success probability will be close to $1$.  We thus ask for groups with trivial center:    
 
\noindent\Question\  1.  As $|G|\rightarrow\infty$, does the probability of solving \HCEP\ approach 1?

 In 1997, Roichman [\Roichman]  studied in greater detail the decomposition of the conjugation representation when $G=S_n$.  He established that (in some precise sense) for most irreducible representations $W$ of $S_n$, the number $m_W$ of copies of $W$ appearing in the conjugation representation is approximately equal to $\dim W$.  Statements of this form, if true for other classes of groups, should force an affirmative answer to Question 1 for those classes of groups.

 \bigskip
\noindent{\bf 8.  Further Questions}

For which groups and which representations can optimal inputs, such as those characterized in Proposition 2, be constructed efficiently? This question applies both to general \SOD\ and to  \GM.   It is easy to see that there is an efficient way to do this in the case of abelian groups. In the case of \GM, Theorem 4 illustrates that one needs only a single isotypic component of $\C^G$ and its dual to produce an optimal algorithm. For groups admitting a nontrivial abelian quotient, this isotypic component may be chosen to be one dimensional and producing an optimal input amounts to identifying these one dimensional subspaces.
  
Another immediate problem is to extend Proposition 4 to give a complete characterization of the inputs that are optimal for \GM. In particular, must we involve those states which are optimal for \SOD\ with respect to the representation of $G \times G$? These states are used in the algorithms provided in Theorems 3 and 4. We took advantage of the fact that such states encode the character of a representation, as expressed in Lemma 3. However it is still unknown whether this is necessary for optimal algorithms solving \GM.

{\bf Acknowledgements}  The authors would like to thank Marcus Robinson, Asif Shakeel, Ethan Edwards and Jon Grice for useful conversations.

\global\setbox1=\hbox{[00]\enspace}
\parindent=\wd1
\noindent{\bf References}
\vskip10pt

\parskip=0pt

\item{[\Kholevo]}
A. S. Kholevo,
``Quantum statistical decision theory'',
\JMA\ {\bf 3} (1973) 337--394.

\item{[\YKL]}
H. P. Yuen, R. S. Kennedy and M. Lax,
``On optimal quantum receivers for digital signal detection'',
\PIEEEL\ {\bf 58} (1970) 1770--1773.

\item{[\Helstrom]}
C. W. Helstrom,
``Detection theory and quantum mechanics'',
\IC\ {\bf 10} (1967) 254--291.

\item{[\HW]}
P. Hausladen and W. K. Wooters, ``A Ôpretty goodÕ measurement for distinguishing quantum states", 
\JMO\ {\bf 41} (1994), 2385--2390.

\item{[\SKIH]}
M. Sasaki, K. Kato, M. Izutsu and O. Hirota,
``Quantum channels showing superadditivity in classical capacity'',
\PRA\ {\bf 58} (1998) 146--158.

\item{[\EldarForney]}
Y. Eldar and G. Forney,
``On quantum detection and the square-root measurement'',
\IEEETIT\ {\bf 47} (2001) 858--872.

\item{[\EMV]}
\emv,
``Optimal detection of symmetric mixed quantum states'',
\IEEETIT\ {\bf 50} (2004) 1198--1207.

\item{[\MeyerPommersheim]}
\dj,
``Single query learning from abelian and non-abelian Hamming distance 
  oracles'',
UCSD preprint (2008).

\item{[\Deutsch]}
D. Deutsch,
``Quantum theory, the Church-Turing principle and the universal 
  quantum computer'',
\PRSLA\ {\bf 400} (1985) 97--117.

\item{[\DeutschJozsa]}
D. Deutsch and R. Jozsa,
``Rapid solution of problems by quantum computation'',
\PRSLA\ {\bf 439} (1992) 553--558.

\item{[\BBCMdW]}
R. Beals, H. Buhrman, R. Cleve, M. Mosca and R. de Wolf,
``Quantum lower bounds by polynomials'',
\JACM\ {\bf 48} (2001) 778--797.

\item{[\MeyerPommersheimB]}
David A. Meyer and James E. Pommersheim,
``Multi-query Quantum Sums"
{\it TQC} 2011: 153--163.

\item{[\vanDam]}
\vandam, 
``Quantum oracle interrogation: Getting all the information for almost half the price'',
{\sl Proceedings of the 39th annual IEEE Symposium on Foundations of Computer Science},
(1998) 362--367.

\item{[\Bhatia]}
R. Bhatia, 
``Positive Definite Matrices'',
Princeton University Press 2007.

\item{[\BV]}
\bv,
``Quantum complexity theory'',
\SIAMJC {\bf 26} (1997) 1411--1473.

\item{[\Frumkin]}
A. Frumkin. ``Theorem about the conjugacy representation of $S_n$",
{\sl Israel J. Math.} {\bf 55.1} (1986): 121--128.

\item{[\Roth]}
Richard L. Roth, ``On the conjugating representation of a finite group",
{\sl Pacific J. Math} {\ bf 36} (1971): 515--521.

\item{[\Formanek]}
E. Formanek. ``The conjugation representation and fusionless extensions", 
{\sl Proc. Amer. Math. Soc.}  {\bf 30.1} (1971): 73--74.

\item{[\Roichman]}
Y. Roichman. ``Decomposition of the conjugacy representation of the symmetric groups", 
{\sl Israel J. Math.}  {\bf 97.1} (1997): 305--316.

\bye
 
\vfill\eject

\item{[\Simon]}
\simon,
``On the power of quantum computation'',
in S. Goldwasser, ed.,
{\sl Proceedings of the 35th Annual Symposium on Foundations of 
     Computer Science}, Santa Fe, NM, 20--22 November 1994
(Los Alamitos, CA:  IEEE 1994) 116--123;\hfb
\simon,
``On the power of quantum computation'',
\SIAMJC\ {\bf  26} (1997) 1474--1483.

\item{[\Shor]}
\shor,
``Algorithms for quantum computation:  discrete logarithms and 
  factoring'',
in S. Goldwasser, ed.,
{\sl Proceedings of the 35th Symposium on Foundations of Computer 
Science}, Santa Fe, NM, 20--22 November 1994
(Los Alamitos, CA:  IEEE Computer Society Press 1994) 124--134;\hfb
\shor,
``Polynomial-time algorithms for prime factorization and discrete 
  logarithms on a quantum computer'',
\SIAMJC\ {\bf 26} (1997) 1484--1509.

\item{[\BonehLipton]}
D. Boneh and R. J. Lipton,
``Quantum cryptanalysis of hidden linear forms'',
in D. Coppersmith, ed.,
{\sl Proceedings of Crypto '95}, 
{\sl Lecture Notes in Computer Science\/} {\bf 963} 
(Berlin:  Springer-Verlag 1995) 424--437.

\item{[\Jozsa]}
R. Jozsa, 
``Quantum algorithms and the Fourier transform',
\PRSLA\ {\bf 454} (1998) 323--337.

\item{[\Angluinsurvey]}
See, \eg,
D. Angluin,
``Computational learning theory:  Survey and selected bibliography'',
in
{\sl Proceedings of the Twenty-Fourth Annual ACM Symposium on Theory
     of Computing\/}
(New York:  ACM 1992) 351--369.

\item{[\Grover]}
\grover,
``A fast quantum mechanical algorithm for database search'',
in {\sl Proceedings of the Twenty-Eighth Annual ACM Symposium on 
  the Theory of Computing},
Philadelphia, PA, 22--24 May 1996 
(New York:  ACM 1996) 212--219;\hfb
\grover, 
``Quantum mechanics helps in searching for a needle in a haystack'', 
\PRL\  {\bf 79} (1997) 325--328.

\item{[\Deutsch]}
D. Deutsch,
``Quantum theory, the Church-Turing principle and the universal 
  quantum computer'',
\PRSLA\ {\bf 400} (1985) 97--117.

\item{[\CEMM]}
R. Cleve, A. Ekert, C. Macchiavello and M. Mosca,
``Quantum algorithms revisited'',
\PRSLA\ {\bf 454} (1998) 339--354.

\item{[\Shamir]}
A. Shamir,
``How to share a secret'',
\CACM\ {\bf 22} (1979) 612--613.

\item{[\BV]}
\bv,
``Quantum complexity theory'',
\SIAMJC {\bf 26} (1997) 1411--1473.

\item{[\MeyerPommersheimB]}
\dj,
``More than useless quantum queries'',
in preparation.

\item{[\HMPPR]}
M. Hunziker, \dajm, J. Park, J. Pommersheim and M. Rothstein,
``The geometry of quantum learning'',
{\tt quant-ph/0309059};
to appear in \QIP.

\item{[\Angluin]}
D. Angluin,
``Queries and concept learning'',
\ML\ {\bf 2} (1988) 319--342.

\item{[\BrassardHoyer]}
G. Brassard and P. H{\o}yer,
``An exact quantum polynomial-time algorithm for Simon's problem'',
{\sl Proceedings of 5th Israeli Symposium on Theory of
      Computing and Systems}, Ramat-Gan, Israel 17--19 June 1997
(Los Alamitos, CA:  IEEE 1997) 12--23.

\item{[\GroverB]}
\grover,
``A framework for fast quantum mechanical algorithms'',
in
{\sl Proceedings of the 30th Annual ACM Symposium on Theory of Computing},
Dallas, TX, 23--26 May 1998
(New York:  ACM 1998) 53--62.

\item{[\BHT]}
G. Brassard, P. H{\o}yer and A. Tapp,
``Quantum counting'',
{\sl Proceedings of the 25th International Colloquium on Automata,
Languages, and Programming}, {\AA}lborg, Denmark, 13--17 July 1998,
{\sl Lecture Notes in Computer Science\/} {\bf 1443}
(Berlin:  Springer-Verlag 1998) 820--831.

\item{[\BHMT]}
G. Brassard, P. H{\o}yer, M. Mosca and A. Tapp,
``Quantum amplitude amplification and estimation'',
in
S. J. Lomonaco, Jr.\ and H. E. Brandt, eds.,
{\sl Quantum Computation and Information},
{\sl Contemporary Mathematics\/} {\bf 305}
(Providence, RI:  AMS 2002) 53--74.

\bye